\newcounter{myctr}

\documentclass{ws-ijqi}

\begin{document}

\markboth{Hans-Thomas Elze}
{Gauge Symmetry of the Third Kind and Quantum Mechanics ... }

\catchline{}{}{}{}{}

\title{THE GAUGE SYMMETRY OF THE THIRD KIND AND \\ 
       QUANTUM MECHANICS AS AN INFRARED LIMIT
}

\author{HANS-THOMAS ELZE
}

\address{Dipartimento di Fisica, Universit\`a di Pisa,\\ Largo Pontecorvo 3, I--56127 Pisa, Italia
\\ elze@df.unipi.it}



\maketitle

\begin{history}
\received{\today}
\end{history}

\begin{abstract}
We introduce functional degrees of freedom  
by a new gauge principle related to the phase of 
the wave functional. Thus, quantum mechanical systems 
are dissipatively embedded into a nonlinear classical dynamical structure. There is a necessary  
fundamental length, besides an  
entropy/area parameter, and standard couplings.  
For states that are sufficiently spread over 
configuration space, quantum field theory is recovered.
\end{abstract}


\section{Introduction}	

We present a new ${\cal U}$(1) gauge symmetry ``of the third kind'', related to 
local gauge transformations in the space of field 
configurations underlying quantum field theory (QFT). 
While results are summarized here, more details may be found in Ref.~\refcite{DGElze}.

Point of departure is the observation 
that the reduction of the potentiality described by the wave function(al) $\Psi$   
to the actuality of the outcome of a measurement process has been left 
outside of quantum theory (in its standard form): \\ 
{\it `` ... , it is an incomplete representation of 
real things, although it is the only one which can be built out of the fundamental 
concepts of force and material points (quantum corrections to classical mechanics). 
The incompleteness of the representation leads necessarily to the statistical 
nature (incompleteness) of the laws.''}~\cite{Einstein} \\ 
In other words, aspects of the theory that concern the concept of 
information have hitherto been left separate from 
the concepts of force and material points, the ``real things'' Einstein refers to. 
Correspondingly, this is reflected in the remarkable derivation of the 
kinematical setting of quantum theory from three information theoretical constraints,    
as discussed, for example, in the recent 
review by Bub:~\cite{Bub} the ``real things'' do not play a role in it.
  
The cut between ``real things'' and ``information about 
real things'' appears to be due to historical contingencies, not unlike the cut between dynamical theories, describing the effects of gravity in particular, and geometrical theories of space and time, until their fusion in general relativity.     

In this note, we propose to reconsider the role of the wave functional. We assume   
that its phase is not only subject to global gauge transformations (``of the first 
kind'') and local ones (``of the second kind''), related to the common field variables 
of QFT: There are also  
functional gauge transformations which are {\it local in the space of field configurations} and 
which attribute a physical `charge' to $\Psi$.   

We recall that local U(1) gauge invariance of the Schr\"odinger equation in quantum mechanics (``first quantization'') is related to the electromagnetic interaction via the classical minimal coupling prescription. Two important aspects should be noted:~\cite{DGElze,Saxon} 
({\bf A}) 
The quantum mechanical model of a charged particle interacting with 
the electromagnetic field descends from   
Maxwell's theory, via minimally coupled classical particle Hamiltonian, 
through its quantization, 
finally to the gauge invariant Schr\"odinger equation. 
Thus, there is a {\it classical} regime where the quantum 
theory is anchored (Copenhagen interpretation). -- 
({\bf B}) 
The Schr\"odinger equation can be also interpreted as    
nonrelativistic limit of the Klein-Gordon equation. However, the latter   
is not acceptable as   
a quantum mechanical single-particle wave equation, since there are    
negative energy states and a `probability' density which is not positive definite. 
Instead, with minimal coupling, the Klein-Gordon 
equation describes   
{\it interacting classical} (!) {\it scalar and electromagnetic fields}.   

An analogous dynamical scheme exists for the functional Schr\"odinger equation 
(``second quantization''), as we shall see. 
It predicts the universal coupling of {\it all} fields that are     
variables in this equation and introduces a fundamental length. 

Guided by our new gauge principle, 
we couple a dynamical functional ${\cal A}$ to the wave functional $\Psi$,  
generalizing the action of 
Dirac's variational principle. Here,   
the would-be-quantum sector described by $\Psi$  
forms a dissipative subsystem of the enlarged structure, effectively 
extending it nonlinearly and nonlocally in the space of field configurations. -- 
Nonlinear extensions of nonrelativistic 
quantum mechanics have been studied earlier by Bialynicki-Birula and Mycielski~\cite{Iwo76} 
and by Weinberg~\cite{Weinberg89}. Works of Kibble~\cite{Kibble} and of Kibble and 
Randjbar-Daemi~\cite{Kibble80} present nonlinear extensions of 
scalar QFT models where parameters 
of the models are quantum state dependent. However, this kind of nonlinearity is not compatible 
with our gauge principle, which dictates a different form of  
{\it nonlinearity} (in $\Psi$) of the wave functional equation.

As we shall see, quantum theory is embedded here into a new framework, where the potentiality represented by $\Psi$ bears the character of a charge. It causes correlations of the underlying fields beyond what is encoded in their usual Lagrangians, with QFT describing an infrared limit. We speculate that a model of this kind might point in the right direction towards unifying the concepts of ``real things'' and ``information about real things''. 

In distinction to other proposals that have quantum mechanics emerge from underlying  
dynamics~\cite{tHdet88,tHdet97,tHdet01,All,E04,I05,Smolin,Adler,tH06}, 
our approach does not  
depend on a particular field theory, such as the Standard Model, or otherwise special 
dynamics. 
We share, however, the tentative conclusion that quantum theory can and should be 
reconstructed as an effective theory describing large-scale  
behavior of {\it fundamentally deterministic degrees of freedom}. 
Quantum states are no longer the primary degrees of freedom. 
Bell's theorem and the predicament 
of local hidden variable theories are circumvented, 
since the implied nonlocality operates at the pre-quantum level.
  
\section{The Gauge Invariant Functional Wave Equation}
The functional Schr\"odinger picture of QFT is convenient for our argument. References to 
related work are given in Ref.~\refcite{DGElze}. -- 
We consider a generic scalar field theory, for simplicity, and refer to 
Ref.~\refcite{DGElze} concerning systems with internal gauge symmetries. 
The functional Schr\"odinger equation is: 
\begin{equation}\label{fSscalar}
i\partial_t\Psi [\varphi ;t]=H[\hat\pi ,\varphi]\Psi [\varphi ;t]\equiv
\int\mbox{d}^3x\Big\{ -\frac{1}{2}\frac{\delta^2}{\delta\varphi^2}
+\frac{1}{2}(\nabla\varphi )^2+V(\varphi )\Big\}\Psi [\varphi ;t]
\;\;, \end{equation} 
with the Hamiltonian $H$ corresponding to the classical Hamiltonian density;  
mass and selfinteraction terms are included in $V(\varphi )$. 
Here, the classical canonical momentum 
conjugate to the field (coordinate) $\varphi$ is substituted by: 
\begin{equation}\label{momentum}
\pi (\vec x)\;\longrightarrow\;\hat\pi (\vec x)\equiv
\frac{1}{i}\frac{\delta}{\delta\varphi (\vec x)}
\;\;, \end{equation}
which implements the field quantization (coordinate representation). 

In analogy to gauge transformations in the 
first quantized Schr\"odinger picture, we now define   
${\cal U}$(1) {\it gauge transformations of the third kind} by:  
\begin{equation}\label{localfuncgt} 
\Psi'[\varphi ;t]\equiv\exp (-i\Lambda [\varphi ;t])\Psi [\varphi ;t]
\;\;, \end{equation} 
where $\Lambda$ denotes a time dependent real functional. 
Such gauge transformations are local 
in the space of field configurations. They differ from the usual    
gauge transformations in QFT. -- 
In fact, the wave equation (\ref{fSscalar}) 
becomes invariant under transformation (\ref{localfuncgt}), if  
new covariant derivatives are introduced:  
\begin{eqnarray}\label{dtcov}
\partial_t&\longrightarrow &{\cal D}_t\equiv
\partial_t+i{\cal A}_t[\varphi ;t]
\;\;, \\ \label{dxcov} 
\frac{\delta}{\delta\varphi (\vec x)}&\longrightarrow &
{\cal D}_{\varphi (\vec x)}\equiv
\frac{\delta}{\delta\varphi (\vec x)}+
i{\cal A}_\varphi [\varphi ;t,\vec x]
\;\;. \end{eqnarray} 
The real functional ${\cal A}$ is 
analogous to the usual vector potential. Generally, ${\cal A}$ depends on $t$;  
it is a {\it functional} of $\varphi$ in Eq.\,(\ref{dtcov}), while it is a 
{\it functional field} in Eq.\,(\ref{dxcov}). 
Distinguishing these components of ${\cal A}$ by subscripts,  
they transform according to: 
\begin{eqnarray}\label{Afunctionalgt} 
{\cal A}'_t[\varphi ;t]&\equiv&{\cal A}_t[\varphi ;t]+
\partial_t\Lambda [\varphi ;t]
\;\;, \\ \label{Afunctiongt}  
{\cal A}'_\varphi [\varphi ;t,\vec x]&\equiv=&
{\cal A}_\varphi [\varphi ;t,\vec x]+
\frac{\delta}{\delta\varphi (\vec x)}\Lambda [\varphi ;t]
\;\;. \end{eqnarray} 
Then, we may also define an invariant `field strength':
\begin{equation}\label{field} 
{\cal F}_{t\varphi}[\varphi ;t,\vec x]\equiv 
\partial_t{\cal A}_\varphi [\varphi ;t,\vec x]
-\frac{\delta}{\delta\varphi (\vec x)}
{\cal A}_t[\varphi ;t]
\;\;, \end{equation} 
i.e., ${\cal F}_{t\varphi}=[{\cal D}_t,{\cal D}_\varphi]/i$. 
      
In order to give further meaning to the coupling between $\Psi$ and ${\cal A}$, 
we have to postulate a consistent dynamics for the latter. 
All elementary fields supposedly are present as the coordinates on which the wave 
functional depends -- presently just a scalar field, besides time. 
We assume the following ${\cal U}$(1) invariant action:  
\begin{equation}\label{Action}
\Gamma\equiv\int\mbox{d}t\mbox{D}\varphi\;\Big\{ 
\Psi^*\Big ({\cal N}(\rho )
\stackrel{\leftrightarrow}{i{\cal D}}_t
-H[\frac{1}{i}{\cal D}_{\varphi},\varphi ]\Big )\Psi 
-\frac{l^2}{2}\int\mbox{d}^3x\;\big ({\cal F}_{t\varphi}
\big )^2\Big\}
\;\;, \end{equation} 
where    
$\Psi^*{\cal N}\stackrel{\leftrightarrow}{i{\cal D}}_t\Psi
\equiv\frac{1}{2}{\cal N}
\{\Psi^*i{\cal D}_t\Psi
+(i{\cal D}_t\Psi )^*\Psi\}$, and with   
a dimensionless real function ${\cal N}$ depending on the density:  
\begin{equation}\label{rho}
\rho [\varphi ;t]\equiv\Psi^*[\varphi ;t]\Psi [\varphi ;t]
\;\;. \end{equation}
We shall see shortly that ${\cal N}$ incorporates a necessary {\it nonlinearity}. 
The fundamental parameter $l$ has dimension 
$[l]=[length]$, for dimensionless measure $\mbox{D}\varphi$ and $\Psi$. 
 
Our action generalizes the 
action for a Schr\"odinger wave functional which has been 
employed for applications of Dirac's 
variational principle to QFT before.~\cite{DGElze,Kibble80}   
It depends on 
$\Psi ,\Psi^*,{\cal A}_t$, and 
${\cal A}_\varphi$ separately. -- Varying $\Gamma$ with respect to $\Psi^*$ (and $\Psi$)
yields the gauge invariant $\Psi$-functional equation of motion (and its adjoint):  
\begin{equation}\label{ginvfSscalar}
\left (\rho {\cal N}(\rho )\right )' 
i{\cal D}_t\Psi [\varphi ;t]
=H[\frac{1}{i}{\cal D}_{\varphi},\varphi ]
\Psi [\varphi ;t]
\;\;, \end{equation}
replacing Eq.\,(\ref{fSscalar}); 
here $f'(\rho )\equiv\mbox{d}f(\rho )/\mbox{d}\rho$. 
Varying with respect to ${\cal A}_\varphi$,  
we obtain: 
\begin{equation}\label{fieldeq} 
\partial_t{\cal F}_{t\varphi}[\varphi;t,\vec x]
=\frac{1}{2il^2}\left ( 
\Psi^*[\varphi;t]{\cal D}_{\varphi (\vec x)}\Psi [\varphi;t]
-\Psi [\varphi;t]({\cal D}_{\varphi (\vec x)}\Psi [\varphi;t])^*
\right )
\;\;. \end{equation}
This invariant `gauge field equation' completes the set of dynamical equations. 

The nonlinear Eq.\,(\ref{ginvfSscalar}) preserves the normalization, i.e. 
$\langle\Psi |\Psi\rangle\equiv\int\mbox{D}\varphi\;\Psi^*\Psi$ is conserved,  
while the overlap of two different states, $\langle\Psi_1|\Psi_2\rangle$, 
generally varies in time. This seems to hint at a probability interpretation, yet 
the continuity equation, Eq.\,(\ref{continuity}) below, shows that this cannot  
be maintained. For ${\cal A}\neq 0$, also the 
{\it homogeneity property} does no longer hold, 
i.e., $\Psi$ and $z\Psi$ ($z\in\mathbf{Z}$) present two different  
physical states~\cite{Iwo76,Weinberg89}. 
This changes essential aspects of the measurement theory~\cite{Kibble} and    
indicates that here QFT is embedded in a {\it classical} framework. --   
Furthermore, the Hamiltonian $H$, unlike in QFT, 
cannot be arbitrarily shifted by a constant $\Delta E$, transforming  
$\Psi\rightarrow\exp (-i\Delta Et)\Psi$. -- Finally, our action is invariant under 
space-time translations and spatial rotations. Elsewhere, we will present $\Gamma$   
in a manifestly Lorentz invariant form, given a suitable background spacetime. 
  
Variation of $\Gamma$ with respect to 
${\cal A}_t$, which is a Lagrange multiplier, yields the  
corresponding gauge invariant `Gauss' law':
\begin{equation}\label{Gauss}
-\int\mbox{d}^3x\;\frac{\delta}{\delta\varphi (\vec x)}{\cal F}_{t\varphi}[\varphi ;t,\vec x]
=\frac{1}{l^2}\Psi^*[\varphi ;t]\Psi [\varphi ;t]{\cal N}(\rho )
\;\;. \end{equation} 
This can be combined with Eq.\,(\ref{fieldeq}) to result in the continuity equation: 
\begin{equation}\label{continuity}
0=\partial_t\Big (\rho{\cal N}(\rho )\Big )
+\frac{1}{2i}\int\mbox{d}^3x\;\frac{\delta}{\delta\varphi (\vec x)}
\left ( 
\Psi^*{\cal D}_{\varphi (\vec x)}\Psi 
-\Psi ({\cal D}_{\varphi (\vec x)}\Psi )^*
\right )
\;, \end{equation}
expressing local ${\cal U}$(1) `charge' conservation in the space of field configurations. --  
The Eq.\,(\ref{continuity}) implies that the 
total `charge' $Q$ has to vanish at all times:\cite{DGElze} 
\begin{equation}\label{charge}
Q(t)\equiv\frac{1}{l^2}\int\mbox{D}\varphi\;\rho{\cal N}(\rho )=0
\;\;. \end{equation}    
Here, the necessity of the nonlinearity becomes obvious. 
Without it, the vanishing total `charge' could not be implemented. 
 
Besides necessarily multiplying the invariant term  
$\Psi^*i{\cal D}_t\Psi$ in $\Gamma$, the nonlinearity is not yet determined. 
A particular choice related to an entropy functional has been studied
in Ref.~\refcite{DGElze}. In this case, one finds that 
the `charge' density $\rho{\cal N}(\rho )$ 
is the deviation of entropy density per unit area  
from a reference density $\rho S/l^2$. Thus, 
also the {\it entropy/area} $S/l^2$  
is a parameter, besides the fundamental {\it length} $l$. (Entropy per area 
is an essential parameter in apparently unrelated work of 
Padmanabhan~\cite{Padmanabhan}, suggesting that 
gravity is intrinsically holographic and quantum mechanical.) In any case,    
the timescale of the $\Psi$-functional evolution shrinks or 
expands in different regions of configuration space, 
depending on the factor $(\rho {\cal N}(\rho ))'$ in Eq.\,(\ref{ginvfSscalar}). 

Two remarks are in order here: ({\bf A}) 
The Eqs.\,(\ref{ginvfSscalar})--(\ref{Gauss}) obey a 
{\it weak superposition principle}~\cite{Iwo76}: The sum of two solutions, $\Psi_{1,2}$, 
that do not overlap, presents also a solution, provided that 
${\cal A}={\cal A}_1+{\cal A}_2$ is determined consistently. -- ({\bf B}) 
As we argued in Ref.~\refcite{DGElze}, our nonlinear extension of QFT is {\it local} 
in the usual sense ({\it microcausality}).\cite{Kibble}  
-- However, suppose we integrated out the `gauge field'.  
The resulting effective equation for $\Psi$ would  
be {\it nonlocal in field space and in space-time}. 

\section{Stationary States, Separability and QFT Limit}
We study the separation of the time dependence in  
Eqs.\,(\ref{ginvfSscalar})--(\ref{Gauss}) with the Ansatz    
$\Psi [\varphi ;t]\equiv\mbox{exp}(-i\omega t)\Psi_\omega [\varphi ]$, 
$\omega\in\mathbf{R}$, and consistently 
assuming {\it time independent} ${\cal A}$-functionals. Thus, 
the Eq.\,(\ref{ginvfSscalar}) yields: 
\begin{equation}\label{PsiZeroEq} 
\left (\rho_\omega {\cal N}(\rho_\omega )\right )' 
\Big (\omega -{\cal A}_t[\varphi ]\Big )\Psi_\omega [\varphi ]
=H[\frac{1}{i}{\cal D}_{\varphi},\varphi ]
\Psi_\omega [\varphi ]
\;\;, \end{equation}
with ${\cal D}_\varphi = \frac{\delta}{\delta\varphi}+i{\cal A}_\varphi$ and  
$\rho_\omega\equiv\Psi_\omega^*[\varphi ]\Psi_\omega [\varphi ]$. 
From Eq.\,(\ref{fieldeq}) follows: 
\begin{equation}\label{fieldeq0} 
\frac{1}{2i}\left ( 
\Psi^*_\omega [\varphi ]{\cal D}_{\varphi (\vec x)}\Psi_\omega [\varphi ]
-\Psi_\omega [\varphi ]({\cal D}_{\varphi (\vec x)}\Psi_\omega [\varphi ])^*
\right )=0
\;\;, \end{equation}
which expresses the vanishing of the `current' in the stationary 
situation. -- 
Applying a time independent gauge transformation, 
cf. Eqs.\,(\ref{localfuncgt}), (\ref{Afunctiongt}), the stationary wave functional 
can be made {\it real}. The Eq.\,(\ref{fieldeq0}) then implies ${\cal A}_\varphi=0$;   
consequently, ${\cal D}_{\varphi}\rightarrow\frac{\delta}{\delta\varphi}$ 
everywhere. Finally, `Gauss' law', Eq.\,(\ref{Gauss}), 
determines ${\cal A}_t$: 
\begin{equation}\label{Gauss0}  
\int\mbox{d}^3x\;\frac{\delta^2}{\delta\varphi (\vec x)^2}{\cal A}_t[\varphi ]
=\frac{1}{l^2}\rho_\omega {\cal N}(\rho_\omega )
\;\;, \end{equation} 
which has to be solved selfconsistently together with Eq.\,(\ref{PsiZeroEq}). -- 
Separation of the time dependence thus leads to two coupled 
equations. One may guess an appropriate time independent 
$\Psi_\omega$-functional. 
Having an action at hand, Eq.\,(\ref{Action}), the parameters of such an Ansatz can be optimized by the variational principle, in analogy       
to Hartree approximation and semiclassical limit of QFT. 
Furthermore, the Eq.\,(\ref{Gauss0}) can be solved formally by functional Fourier 
transformation, eliminating  ${\cal A}_t$ at the 
expense of introducing the nonlocality mentioned before. 

Turning to {\it separability}, this is an important property of linear 
quantum theory. It allows to combine subsystems which do not interact with each other, without 
creating unphysical correlations~\cite{Iwo76,Weinberg89}. This should be preserved to the  
extent that is verified experimentally. -- We have argued that linear {\it QFT 
arises in the infrared (IR) limit}, and consequently separability:~\cite{DGElze}   
Assume that the system is in a  
{\it diffuse state}, characterized by a density $\rho_\omega$ that is widely spread 
over the space of configurations of $\varphi$. For such a high entropy state, 
the local energy density and the `charge density' on the 
right-hand side of Eq.\,(\ref{Gauss0}) must be small. For our particular 
choice of nonlinearity, then, the stationary functional Schr\"odinger equation results, $\omega\Psi_\omega =H\Psi_\omega$,  
and with it the known structures of QFT. -- 
Consequences of small violations of this linear equation, related to terms 
involving $\rho_\omega$ or ${\cal A}_t$, should be explored. Nonlinear effects become important in our framework only for states with a small uncertainty in configuration space, such that the IR limit does not apply.  

Finally, we remark that two stationary solutions, $\Psi_{\omega_{1,2}}$, of the present 
eigenvalue problem, in general, obey a generalized {\it orthogonality} relation that 
reduces to the usual one of QFT in the IR limit.~\cite{DGElze}  

\section{Conclusions}
New gauge transformations ``of the third kind'' attribute a ${\cal U}$(1) `charge' to the wave functional, which leads to an embedding of 
quantum field theory in a larger nonlinear structure. -- It differs from  
earlier proposals of nonlinear generalizations of quantum mechanics or  
QFT~\cite{Iwo76,Weinberg89,Kibble,Kibble80}.  
We tentatively interpret it 
as a classical one, since differently charged components of the wave functional $\Psi$, 
besides being governed by the usual interactions 
of an underlying field theory model (such as gauge theories, see Ref.~\refcite{DGElze}), 
are coupled through a new connection functional 
${\cal A}$. When effects of the latter are negligible, QFT is recovered.     
  
A number of interesting problems need further study, before this proposal 
can stand on its own. -- A theory of the observables and the measurement process needs to be worked out. It is promising that the energy-momentum tensor following from our action, 
Eq.\,(\ref{Action}),   
similarly in the presence of internal gauge symmetries, is the one of the 
respective QFT {\it plus} contributions due to the coupling between 
$\Psi$ and ${\cal A}$. When the latter is small, the usual observables may be useful, 
while the coupling might be important for the reduction or collapse of the wave 
functional. -- A solution in the case of an underlying free field theory  
should be possible, based on the variational principle, for example. This will be helpful 
to better understand effects of the new coupling. -- 
Reparametrization invariant models are an important target. As compared 
to a Wheeler-DeWitt type equation, the presence of  
additional nonlinear terms in what replaces this equation may actually be useful.       

\section*{Acknowledgments}

I wish to thank G.\,'t\,Hooft for discussions, 
particularly concerning the phase of the wave function.~\cite{tH06}  
I am grateful to M.\,Genovese for the kind invitation 
to the Torino workshop ``Advances in Foundations of Quantum Mechanics and 
Quantum Information with Atoms and Photons''.

\end{document}